\documentclass[fleqn]{annalen}
\usepackage{graphics} 
\begin{document}
\newcommand{\volume}{8} 
\newcommand{\xyear}{1999} 
\newcommand{\issue}{5}
\newcommand{\recdate}{29 July 1999}
\newcommand{\revdate}{dd.mm.yyyy} 
\newcommand{\revnum}{0}
\newcommand{\accdate}{dd.mm.yyyy} 
\newcommand{\coeditor}{ue} 
\newcommand{\firstpage}{507}
\newcommand{\lastpage}{510} 
\setcounter{page}{\firstpage}
\newcommand{\keywords}{weak
localization, mesoscopic conductors,
current noise}
\newcommand{\PACS}{61.43.--j, 71.30.+h,
73.40.Hm}
\newcommand{\shorttitle}{Yuli V.
Nazarov, Universality's of weak
localization } 
\title{Universalities of weak
localization}
\author{Yuli V. Nazarov}
\newcommand{\address} {Faculty of
Applied Physics and DIMES, \\ Delft
University of Technology, \\ Lorentzweg
1, 2628 CJ Delft,The Netherlands}
\newcommand{\email}{\tt
yuli@duttnto.tn.tudelft.nl} \maketitle
\begin{abstract} We present here a novel
approach to evaluation of the weak
localization correction (WLC) to
transport properties of a mesoscopic
conductor. It is based on an extension
of Keldysh technique and allows one to
evaluate the full counting statistics of
the current in the conductor. In our
opinion, it provides a fresh look on the
theory of weak localization.
\end{abstract}

\section{Introduction and method}
\label{points}

Since the discovery of weak localization twenty years ago
\cite{Four} much attention has been given to its scaling
properties in low dimensions in infinite homogenious media. 
There is also weak localization correction to transort properties of a
mesoscopic conductor that is shorter than decoherence
length. Such a conductor is neither scalable nor
homogeneous, so that one generally expects the WLC to depend
on details of sample structure. Still, the WLC exhibits some
universal features.

We concentrate in this work on evaluation of WLC to transport
properties of two-terminal mesoscopic diffusive conductor.
We employ here a new technique that allows to calculate not
only the current in the conductor but also all noise
characteristics of the current, so-called {\it full counting
statistics}.\cite{LLL} Beside the fact that this method
provides more information about transport properties of the
conductor than the traditional Green's function \cite{Four}
and RMT \cite{Carlo} methods, it gives compact and clear
general expressions that are valid for inhomogeneous
conductors.

Due to the lack of space, it is impossible to review all
the results obtained. We concentrate on the most important
ones that concern the universal features of WLC: its form
for quasi-one-dimensional conductors, general formula for the
current correction and its possible dependence on momentum relaxation time, 
universal Cooperon mode. We disregard influence of magnetic
field and spin-orbit interaction of the WLC.

The method in use presents
an extension of Keldysh Green function (KGF) method (see
\cite{Rammer} for review) and is based on the following
trick: let us define a one-electron Green function by means
of 
\begin{equation} ( i\frac{\partial}{\partial t} - \check
H - \tau_3 \chi(t)\check I ) \otimes \check G= \delta(1-1')
\end{equation} 
(notations of Ref. \cite{Rammer}), where
$\check I$ is for the time being an one-electron arbitrary
operator, $\chi(t)$ is a time-dependent parameter, $\tau_3$
is $2 \times 2$ matrix in Keldysh indexes. It is easy to
show by traditional diagrammatic methods that the expansion
of $\check G$ in $\chi(t)$ generates diagrams for higher order
correlators of $\check I$. 

The trick can be readily applied to the problem of full counting statistics
of the current.\cite{LLL} To this end, we set $\check I$
to the operator of the current through a cross section of
the conductor, $\chi(t)$ to constant value. It is convenient
to introduce the following $\chi$-dependent action to express the probability for 
$N$ electrons to be transferred 
through the conductor during time interval $t$
\begin{equation}
\frac{\partial S}{\partial \chi} = 
i t \int \frac{d\varepsilon}{2\pi}{\rm Tr} \ \big( \tau_3 \check I  \check 
G(\varepsilon) \big); / / /
P(N)= \int_{-\pi}^{\pi} \frac{d \chi}{2 \pi} e^{-S(\chi) -i N \chi}.
\end{equation}
Derivatives of $S$ give moments of $P(N)$, first and second
derivative corresponding to average current and current noise
respectively.

\subsection{Semiclassical limit}
\label{semic}
To calcultate $S$ in semiclassical limit for duffusive
conductor, one  derives a diffusion equation for
the new KGF in coinciding points, 
$\check G(x,x) \equiv \pi \nu \hat G(x)$, $\nu$ being density of
states. It turns out that $\hat G^2=1$ and the equation takes
traditional form \cite{Larkin, Nazarov}
\begin{equation}
\frac{\partial \hat j_\alpha}{\partial x_\alpha}=0; 
\ \ \hat j_\alpha = - {\cal D}(x) \hat G \frac{\partial }{\partial x_\alpha} \hat G.
\label{difu}
\end{equation}
at any energy, {\cal D} being diffusive coefficient. The boundary conditions to Eq. 
\ref{difu} are set by KGF in terminals,
terms with $\chi$ can be incorporated into these conditions, so that $\hat G$ equals
\begin{equation}
\hat G_1 = \left(\matrix{ 1- 2 f_1 & 2 f_1 \cr 2(1-f_1) & 2f_1 -1}\right) ;
\hat G_2 = \left(\matrix{ 1- 2 f_2 & 2 f_2 e^{i \chi} \cr 2(1-f_2) e^{-i \chi} & 2f_2 
-1}\right). 
\end{equation} 
in the right and left reservoir respectively, $f_{1,2}(\varepsilon)$ being
electron distribution functions in corresponding reservoirs.
The equation \ref{difu} can be solved in a rather general form yielding
\begin{equation}
\check j_\alpha = - {\cal D}(x) \ln(\hat G_1 \hat G_2) \frac{\partial u(x)}{\partial 
x_\alpha} .
\end{equation}
where $u(x)$ satisfies diffusion equation
with boundary conditions $u_1=0, u_2=1$. Physically, $u(x)$
is a normalized voltage distribution in the conductor. This
solution leads to the following quasiclassical action:
\begin{eqnarray}
-S_{class} = \frac{t G}{4 e^2} \int d \varepsilon M(\varepsilon);\\
M \equiv \ln^2 \big[ (2f_1-1)(2f_2-1) + 2(f_1(1-f_2)e^{-i\chi}+f_2(1-f_1)e^{i 
\chi})\\
+2 \sqrt{2 i \sin \frac{\chi}{2} (f_1 e^{i \chi}+1-f_1) (f_2 e^{-i \chi}+1-f_2)
(f_2(1-f_1)e^{-i \chi/2} -f_1(1-f_2)e^{i \chi/2})}\big]
\end{eqnarray}
$G$ being full conductance. 
This generalizes results of \cite{LeeLev}, those were obtained
by RMT methods. 
One can find in \cite{Nazarov} why both methods give the same
results. 

\subsection{Weak localization correction}
\label{correction}

\begin{figure}
\centerline{\includegraphics*[6cm,4cm]{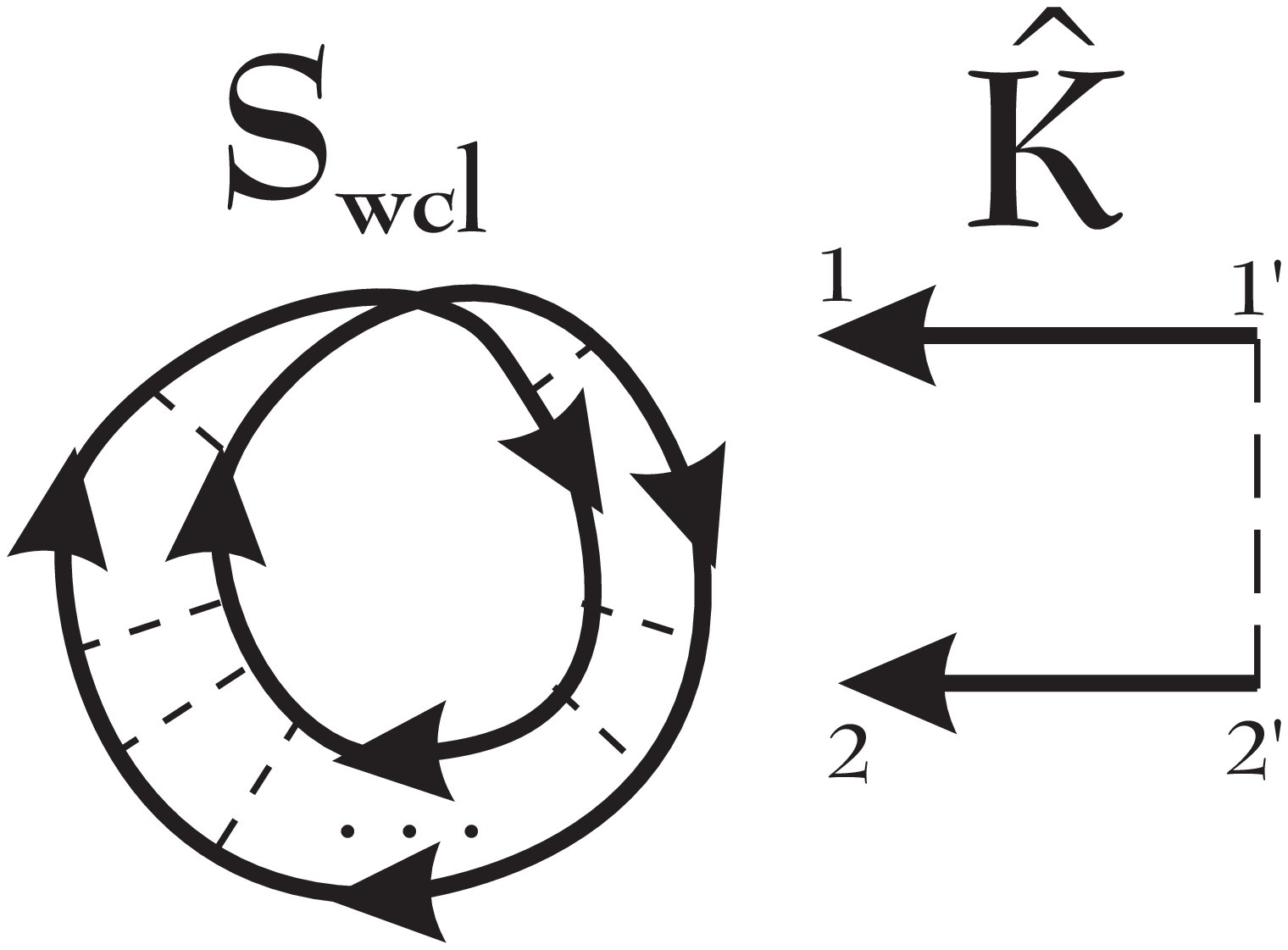}}
\caption{}
  \label{sample_ps_fig}
\end{figure}

The advantage of the method is that one obtains the WLC directly to $S$. It
is given by a usual Cooperon ladder \cite{Four} (Fig. 1) made of new KGF. Due
to its ladder nature, the correction can be express in terms of eigenvalues
 $1+ \lambda_n$ of the ladder section operator 
\begin{equation}
K^{ab,cd}(1,2;1',2') = G^{ac}(1,1')U(1'-2')G^{bd}(2,2')
\end{equation}
where $U(1'-2')$ presents impurity potential. It reads
\begin{equation}
-S_{wlc} = t \int \frac{d \varepsilon}{\pi} \sum_n \ln\lambda_n
\label{wcl}
\end{equation}
Here dimensionless $\lambda_n$ characterize Cooperon propagation.
We derive corresponding diffusion equation for these eigenvalues
and matrix eigenfunction $\hat f$,
\begin{equation}
-\frac{\partial}{\partial x_\alpha} {\cal D}(x) 
\big( \hat G  \frac{\partial}{\partial x_\alpha} \hat f+
\hat f \frac{\partial}{\partial x_\alpha} \hat G^T \big)
= \lambda ( \hat G \hat f - \hat f \hat G^T) \tau(x)
\end{equation} 
It is essential that the eigenvalues do 
depend on $\tau(x)$, the momentum relaxation time. This is not
the same as transport scattering time that enters diffusion
coefficient and in principle shall be considered as an independent
parameter. The relevant branch of eigenvalues corresponds to 
$\hat f \propto  \tau_2$. That simplifies the equation
considerably. Finally we arrive at the following relation for
eigenvalues $\lambda_n$
\begin{equation}
-\tau^{1/2}(x)\frac{\partial}{\partial x_\alpha} {\cal D}(x)
\frac{\partial}{\partial x_\alpha} \tau^{1/2}(x)f
- {\cal D}(x) \tau(x) M(\varepsilon) (\nabla_\alpha u)^2 f = \lambda f
\end{equation}
that implicitly presents the result for the WLC in the most general form.  
\section{Results and discussion}
\subsection{Universality in  a general conductor}
We note an interesting feature of the results: although each eigenvalue
$\lambda$ does depend on $\tau(x)$, the WLC does not since the corrections
to different $\lambda$ cancel each other. So that we can replace $\lambda$ by 
effective $\tilde \lambda$ those are given by equation that does not contain 
$\tau(x)$
\begin{equation}
-\frac{\partial}{\partial x_\alpha} {\cal D}(x)
\frac{\partial}{\partial x_\alpha} f
- {\cal D}(x)  M(\varepsilon) (\nabla_\alpha u)^2 f = \tilde\lambda f
\end{equation}

Since $M \approx i 4 \chi (f_1-f_2)$ at $\chi \rightarrow 0$,
and the average current $ \propto \partial S/\partial \chi(0)$,
the WLC to conductivity is obtained by perturbation expansion
in $M$ yielding
\begin{equation}
 G_{wlc} = - \frac{4 e^2}{\pi} \sum_n \frac{<{\cal D}(x)
(\nabla_\alpha u)^2>_n}{\tilde\lambda_n(M=0)}
\end{equation}
where brackets denote average with respect to $n$-th
Cooperon state at $M=0$.

\subsection{Universality in a quasi-one-dimensional conductor}
There is even more universality  in the case of quasi-one-dimensional
conductor where {\cal D}(x) depends only on single coordinate and only one 
Cooperon
branch has to be taken into account. The eigenvalues reduce
to $\pi^2 n + M$, $n \ge 1$ and the spatial dependence of diffusion 
coefficient is of no importance.
The sum in \ref{wcl} can be taken explicitly yielding
\begin{equation}
S_{wcl} = t \int \frac{d \varepsilon}{\pi} \ln \big(\frac{\sin(\sqrt{ 
-M})}{\sqrt{-M}} \big)
\end{equation}
\subsection{Universal gapless mode}
At $\chi=0$ Cooperon eigenvalues never approach zero. This
is a kind of finite-size effect specific for mesoscopic
conductor. We note that there is a universal Cooperon mode
with eigenfunction $\sin(\pi u(x))$ that becomes gapless
in the absence of magnetic field provided $M \rightarrow -\pi^2$. 
This indicates a strong divergence of WLC in this regime 
that can be lifted by magnetic field. $M$ approaches $-\pi^2$ 
in the limit of $\chi \rightarrow i \infty, f_1=0, f_2=1$
that corresponds to a current fluctuation which is much smaller than
the average current. These fuctuations are exponentially 
suppessed by $S_{class}$, and diverging $S_{wlc}$ provides
anomalous magnetic dependence of their probability.

\vspace*{0.25cm} \baselineskip=10pt{\small \noindent I am indebted to
C. J. W. Beenakker, D.E. Khmelnitski and L. S. Levitov for
several enlighting discussions. }
%
%
%
%
%
%
%
%
%
%
%
%

\end{document}